\documentstyle[aps,pra,epsf]{revtex}
\begin{document}
\draft

\newcommand{\pp}[1]{\phantom{#1}}
\newcommand{\be}{\begin{eqnarray}}
\newcommand{\ee}{\end{eqnarray}}
\newcommand{\ve}{\varepsilon}
\newcommand{\vs}{\varsigma}
\newcommand{\Tr}{{\,\rm Tr\,}}
\newcommand{\pol}{\frac{1}{2}}
\newcommand{\ba}{\begin{array}}
\newcommand{\ea}{\end{array}}
\newcommand{\bear}{\begin{eqnarray}}
\newcommand{\eear}{\end{eqnarray}}
\title{
Schr\"odinger-picture correlation functions for nonlinear evolutions 
}
\author{Marek~Czachor$^{1,2}$ and Heinz-Dietrich Doebner$^{2}$}
\address{
$^1$ Katedra Fizyki Teoretycznej i Metod Matematycznych\\
Politechnika Gda\'{n}ska,
ul. Narutowicza 11/12, 80-952 Gda\'{n}sk, Poland\\
$^2$ Department of Physics, Technische Universit\"at Clausthal\\
38678 Clausthal-Zellerfeld, Germany}
\maketitle

\begin{abstract}
The well known interpretational difficulties with nonlinear
Schr\"odinger and von Neumann equations can be reduced to the problem
of computing multiple-time correlation functions in the absence of
Heisenberg picture. Having no Heisenberg picture one often resorts to
Zeno-type reasoning which explicitly involves the projection
postulate as a means of computing conditional and joint
probabilities. Although the method works well in linear quantum
mechanics, it completely fails for nonlinear evolutions. We propose
an alternative way of performing the same task in linear quantum
mechanics and show that the method smoothly extends to the nonlinear
domain. The trick is to use appropriate time-dependent Hamiltonians
which involve ``switching-off functions".
We apply the technique to the EPR problem in nonlinear quantum
mechanics and show that paradoxes of Gisin and
Polchinski disappear.
\end{abstract}


\section{Introduction}

Ten years ago Polchinski published a paper \cite{Polchinski}
where he showed how to
avoid unphysical influences between separated systems in nonlinear
quantum mechanics. The paper was a response to criticisms  of
Kibble-Weinberg nonlinear quantum mechanics \cite{Kibble,Weinberg} 
raised by Gisin \cite{Gisin}
and one of us \cite{MCu}. Polchinski proposed to
eliminate the problem by eliminating its source, the projection
postulate, and resorting to the Many Worlds Interpretation. 
However, after having eliminated one difficulty he showed
that there appears another one, a kind of weird communication between
different Everett worlds. 

The paper had several implications. First of all, it convinced a
large group of physicists working on foundations of quantum
mechanics, including Weinberg himself (see his remarks in \cite{Dreams}),
that linearity is an ultimate element of all the possible ``final
theories". 

On the other hand, the Polchinski multiparticle extension
showed that at least part of difficultes trivally disappears if one
starts with nonlinear evolutions formulated in terms of von
Neumann-type equations. Looking at the von Neumann dynamics as a
classical Hamiltonian system on a Poisson manifold of states
represented by reduced density matrices, one arrives at a
surprisingly elegant formalism 
\cite{Bona,Jordan,MC96,MC98,MC97,MCMK98} 
which goes much beyond the original Weinberg
proposal. 
Theories so constructed led to new families of integrable
von Neumann equations which turned out to be very interesting in
themselves at least from the point of view of integrable systems
\cite{SLMC98,MKMCSL99,NUMCMKSL00,NUMC00,JC01}. 

In spite of the original problems with the Weinberg
formalism notable developments were reported by groups working on ``geometric
quantum mechanics" \cite{Cirelli,Ashtekar,Hughston}. Quantum mechanics, when
looked at from a geometric standpoint, turns out to be a classical
theory ``living" on a very symmetric phase space (a K\"ahler
manifold with maximal symmetries). It becomes quite natural, then, to
think of standard quantum mechanics as an analogue of a ``vacuum"
spacetime from general relativity. Since the most interesting
spacetimes are those which are less symmetric, maybe the same will be
true for quantum mechanics? An unquestionable beauty of the
geometric perspective suggests that these possibilities should be
further investigated.  

Yet another road to nonlinear Schr\"odinger equations
was discovered by Goldin and Doebner \cite{DG,DGN,Gnlmp,Mizrahi}. 
Studying representations of groups of
diffeomorphisms they noticed that there exists a class of
representations parametrized by the real number $D$ which has an
interpretation of a diffusion coefficient \cite{DG1,DG2}. 
All the other representations had a clear physical interpretation. 
They were known to unify the
description of an extraordinary variety of quantum systems and
generated some physical predictions difficult to reach by other means
(e.g. the anyon statistics in two-dimensional space). 
However, this particular class resisted a quantum mechanical
interpretation until it was realized that it corresponds to a family
of nonlinear dissipative Schr\"odinger equations earlier derived in the
context of quantum chemistry by Schuch and his collaborators
\cite{Schuch}. Doebner and Goldin subsequently introduced the group
of nonlinear gauge transformations necessary to understand the
resulting theory. The resulting formalism was shown to contain as
special cases nonlinear modifications of quantum mechanics proposed
independently by other researchers
\cite{Kostin,BBM,HB,Guerra,Stenflo,Sabatier,Malomed}
It became clear
that linearity is a gauge-dependent property and a simple rule
``linear good, nonlinear bad" is oversimplified. 

Quite recently it turned out that dynamics of certain
$D$-branes leads to nonlinear Schr\"odinger equations od the
Doebner-Goldin type
\cite{Mavromatos1,Mavromatos2} and the von Neumann-type nonlinearities 
introduced in \cite{MC97} 
have links with nonextensive statistics \cite{MCJN99,T88,TMP98}.  

In spite of the optimism of the groups we have mentioned not everyone
was fully satisfied by the Many Worlds solution. Gisin
openly disqualified nonlinear Schr\"odinger equations as
``irrelevant" \cite{Gisin-irr}. Goldin \cite{Goldin} did not agree
with Gisin but formulated a series of doubts with respect to the
Polchinski solution pointing out, in particular, that multiparticle
extensions of physically equivalent theories but performed in different
nonlinear gauges lead to physically inequivalent results. 
Also Mielnik in a recent paper \cite{Mielnik00} argued on general
grounds that the difficulties remain. 

As a result, a person who wants to understand the current state of
knowledge about nonlinear generalizations of quantum
mechanics will find a complete spectrum of mutually
contradicting statements. 

We think that three main questions are causing the entire embarassment.
One is the unclear role of nonlinear averages and 
eigenvalues if nonlinear observables are involved.  The other is 
the role of density matrices (``mixed" states) in nonlinear formalism. 
Finally, an open question is how
to compute joint probabilities in EPR-type experiments.

The main topic of the present paper is the third issue. Two
applications of nonlinear averages are discussed in the Appendix. 
We do not address the problem of density matrices. We believe that 
a common error one finds in the literature on nonlinear Schr\"odinger
equations is to treat solutions of von Neumann-type equations as
states which are {\it mixed\/} in the standard sense of mixed
states on a phase space \cite{Bona1}. To the best of our knowledge no
approach based on Schr\"odinger-type equations was able to
successfully address the question of composite systems. This should
be contrasted with the formalism based on von Neumann equations
which, up 
to a few interpretational subtleties which are a subject of the
present work, simply does not lead to this difficulty. 

In what follows we formulate a new operational
framework for interpretation of experiments in both linear and
nonlinear quantum systems. We do not use the Many Worlds
Interpretation and hence do not have the ``Everett phone".
On the other hand we make a very restrictive use of the projection
postulate. 
We allow to use projections only in two cases: (1) to determine
initial conditions and (2) at the moments particles are 
detected and destroyed (final conditions). 

The question that arises, then, is how to describe situations where
there are, say, two correlated particles and one of them is destroyed
by a detector at time $t=t_1$ whereas the other one is allowed to evolve
until it gets destroyed at $t=t_2>t_1$. This is what we call a
two-time measurement. A typical solution, known
also as a quantum Zeno effect, is to regard the second particle as
evolving for $t_1<t<t_2$ with a new initial condition produced by an
appropriate projection-at-a-distance at $t=t_1$. 

We exclude this approach since in application to nonlinear
evolutions of entangled states it leads to serious complications. Instead, we
reformulate the standard strategy with the help of 
``switching-off functions" and show that (1) in the linear case the result is
identical to the standard Zeno-type calculation and (2) in nonlinear
case is free from unphysical influences between correlated particles
and, hence, ultimately eliminates the ``EPR phone". 

We will begin with a few remarks on filters and histories. The notion
of a filter often occurs in the context of conditional probabilities
and the projection postulate. However, there exists also a desription
where filters are described in terms of unitary matrices and do not
lead to irreversible projections. This point is essential for our
argument. In the next two sections we discuss the role of the projection
postulate for two-time measurements and point out the role of
``switching-off functions". In Sec. V we discuss problems with two-time
measurements in presence of nonlinearities. Sec. VI discusses
Polchinski's multi-particle extension and its modification in terms
of switching-off functions. In Sec. VII we explicitly solve an example and
compare the standard projection-based calculation with the new
proposal. Sec. VIII discusses implications of our approach for the
question of teleportation. Arguments are given that one should not
identify pre- and post-selected ensembles if nonlinear evolutions are
involved. The point is essential since a true teleportation
corresponds to a pre-selected ensemble whereas ordinary correlation
experiments involve post-selections.

We end with the Appendix on nonlinear averages,
the use of them in information theory and nonextensive statistics, and their
links with the subject of the paper. 

\section{Histories and filters}

A typical quantum history is shown in Fig. 1. A particle produced by
a source undergoes a series of scattering events, represented by
unitary evolutions $U_1$ and $U_2$. Before, between and after the
scattering the evolution is free, say. The particle may, for
simplicity, be regarded as a two level system and the $U$s as some
sorts of beam splitters (mirrors, analyzers of polarizaton,
Stern-Gerlach magnets,...). There are several outgoing channels, and
each of the channels corresponds to a different result of the experiment. 

The probability of 
detection in, say, channel 1 can be calculated in a couple of
different ways. One way of doing this is to compute
the overall evolution operator $V(t)$, including the whole
evolution from the source till detection, and calculate the
probability 
\be
p_1
&=&
\Tr\big( E_2 V(t)\rho(0)V(t)^{\dag} \big)\nonumber\\
&=&
\Tr\big(E_2 \rho(t)\big)
\label{h'}
\ee
where $E_2$ is an appropriate projector operator. 

Another way is to split the evolution into pieces explicitly
involving the scattering matrices $U_k$ and a free evolution between
them. The $U$s play here a role of analyzers of the properties
represented by projection operators $E_k$ and ${^\perp E}_k$,
$E_k+{^\perp E}_k=I$. Denote by  $U(t-t')$ the free evolution
between times $t$ and $t'$. Since it satisfies the group composition
property 
\be
U(t-t')=U(t')^{\dag}U(t).\label{group}
\ee
we find  
\be
p_1
&=&
\Tr\Big(U(t-t_2)E_2U(t_2-t_1) E_1 U(t_1)
\rho(0)U(t_1)^{\dag}E_1U(t_2-t_1)^{\dag}E_2U(t-t_2)^{\dag} \Big)
\label{h'1}
\\
&=&
\Tr\Big(E_2(t_2)E_1(t_1)
\rho(0)E_1(t_1)E_2(t_2)\Big)\label{h'2}
\ee
where $E_k(t_k)=U(t_k)^{\dag}E_kU(t_k)$. Expression (\ref{h'2}) is
often called a {\it history\/} \cite{Omnes}. In quantum-optics
literature one would rather speak of a two-time correlation function.
In derivation of (\ref{h'1}) one 
makes an explicit use
of the unitarity of evolution whereas the transition from (\ref{h'1}) to
(\ref{h'2}) involves (\ref{group}). 
On the other hand, the form (\ref{h'}) is more general and can
survive even if the dynamics $t\mapsto \rho(t)$ is nonlinear. 

It is an appropriate moment to point out that 
one should not treat the first splitting of the beam by
$U_1$ as a {\it measurement\/} of the property $E_1$. The measurement
occurs when the particle vanishes in the detector and produces a
click. Before that moment the evolution is unitary and reversible. 
After the measurement the measured particle is destroyed and is no
longer under the jurisdiction of quantum mechanics. 

Our beam splitters $U_k$ are examples of devices which are sometimes
referred to as 
{\it filters\/} \cite{Mielnik-filters}.
As we have seen, the derivation of the history
(\ref{h'2}) explicitly employed the fact that, for calculational purposes, 
filters can be represented by projectors and, hence, by non-unitary
operators. However, it is extremely
important to understand that the same physical process can be
described by the exact formula (\ref{h'}) which describes propagation
of particles through filters in a {\it unitary\/} way. 

In such a formulation projectors are applied only twice --- at the
very beginning when one selects a concrete initial condition, and at the very
end where the particle is detected and destroyed. 

\section{Entangled states and two-time measurements}

Two-time measurements are at the heart of difficulties with nonlinear
extensions of quantum mechanics. We begin with their orthodox
description in linear quantum mechanics. 

Consider a two-electron system initially prepared in the
singlet state 
\be
|\Psi(0)\rangle &=& 
\frac{1}{\sqrt{2}}\Big(|+\rangle |-\rangle
-|-\rangle |+\rangle
\Big)
\ee
and whose Hamiltonian is 
\be
H&=& H_1\otimes I_2+I_1\otimes H_2.
\ee
The dynamics of the state is 
\be
|\Psi(t)\rangle &=& 
V_1(t)\otimes V_2(t)|\Psi(0)\rangle.
\ee
where $V_k(t)=\exp (-i H_kt)$. 
It is known that $|\Psi(0)\rangle$ will
be unchanged if $V_1=V_2$. However, if $H_1\neq H_2$ the perfect
initial anti-correlation of spins holding at $t=0$ will be lost at
later times.  

Now let us imagine that two measurements of the $z$-components of
spin are performed; one 
at time $t=t_1$ on particle $\#1$, and the other at $t=t_2>t_1$ on
particle $\#2$. 
We want to calculate an average of the observable $\sigma_z\otimes
\sigma_z$. 

Denoting by $E_\pm\otimes I$ and $I\otimes E_\pm$ the appropriate
projectors we can proceed as follows. 
\begin{itemize}
\item
At $t=t_1$ project 
\be
|\Psi(t_1)\rangle\mapsto \frac{E_\pm\otimes I |\Psi(t_1)\rangle}
{\parallel E_\pm\otimes I |\Psi(t_1)\rangle\parallel}\label{step1}
\ee
\item
Evolve the resulting state by
\be
\frac{E_\pm\otimes V_2(t_2-t_1)
|\Psi(t_1)\rangle} 
{\parallel E_\pm\otimes I |\Psi(t_1)\rangle\parallel}\label{step2}
\ee
\item
Calculate at $t=t_2$ the (conditional) probabilities 
\be
\frac{\langle \Psi(t_1)|
E_\pm\otimes V_2(t_2-t_1)^{\dag}E_{\pm'}V_2(t_2-t_1) |\Psi(t_1)\rangle}
{\langle \Psi(t_1)|E_\pm\otimes I |\Psi(t_1)\rangle}\label{step3}
\ee
\item
The denominator in (\ref{step3}) is the probability of finding $\pm
1$ for the first particle at $t=t_1$. Therefore
the joint probability of finding $\pm$ for the first particle and
$\pm'$ for the second one is the numerator of (\ref{step3}), i.e.
\be
\langle \Psi(0)|
V_1(t_1)^{\dag}E_{\pm}V_1(t_1)\otimes
V_2(t_2)^{\dag}E_{\pm'}V_2(t_2) |\Psi(0)\rangle.\label{step4}
\ee
\end{itemize}
The second step (\ref{step2}) was done in this way since we knew
already the result 
of one measurement after (\ref{step1}). 
The result created an initial condition for
the remaining dynamics of the other particle at $t=t_1$ (this is
essentially the quantum Zeno effect). The projection
postulate has therefore explicitly been used in the second step at
$t=t_1$. 

The formula (\ref{step4}) is again a correlation function. 
Let us note that we would get the same formula if we
computed the average of the projector
\be
E_{\pm}\otimes E_{\pm'}\label{proj}
\ee
in the state 
\be
V_1(t_1)\otimes V_2(t_2)|\Psi(0)\rangle.\label{frozen}
\ee
An identical result would have been obtained if one considered the
time dependent Hamiltonian 
\be
H_{t_1,t_2}(t)&=& \theta(t-t_1) H_1\otimes I_2+
\theta(t-t_2) I_1\otimes H_2.\label{kill}
\ee
were the $\theta(x)$s is the step function equal 1 for $x<0$ and 0
otherwise, and computed the average of (\ref{proj}) in the state
\be
|\Psi_{t_1,t_2}(t)\rangle
=
e^{-i H_1\otimes I_2\int_0^t \theta(\tau-t_1)d\tau   
-iI_1\otimes H_2\int_0^t\theta(\tau-t_2)d\tau}
|\Psi(0)\rangle.\label{state12}
\ee
The times $t_1$ and $t_2$ play a role of parameters describing the system. 
They naturally incorporate into the formalism the presence of
detectors and the fact that detections are registered at certain
times. 

It is easy to understand why the energy of the particles is not
conserved during the dynamics: At the moments the particles ``get
frozen" by the detection process their energies are transferred to
the detectors. Only the
composite system ``particles plus detectors" is closed. 

It is also obvious that one should not apply ``switching-off
functions" $\theta$ for acts of filtering that do not end up in a
detection and destruction, such as those denoted by $U_1$ and $E_1$ 
at Fig. 1. 

\section{Two-time measurements on beams of pairs}

In real experiments one deals with beams of pairs. A beam consisting
of $N$ pairs is a tensor product of $N$ copies of single-pair states.
Real experiments do not produce all the pairs at the same initial
$t=0$. An experiment may run for weeks, therefore it is much more
realistic to impose initial conditions on each of the pairs
separetely and at different times: $t_0^1,\dots ,t_0^N$. 

A run consisting of $N$ pairs of detections at times
$(t_1^1,t_2^1)$, ... $(t_1^N,t_2^N)$, can be described by
the state vector
\be
|\Psi_{\{t_0^i,t_1^i,t_2^i\}}(t)\rangle
&=&
|\Psi_{t_0^1,t_1^1,t_2^1}(t)\rangle
\otimes 
\dots
\otimes
|\Psi_{t_0^N,t_1^N,t_2^N}(t)\rangle.\label{N pairs}
\ee
If experimental conditions do not change during the run one may
parametrize (\ref{N pairs}) by times of flight $\Delta
t_k^i=t_k^i-t_0^i$. 

Observables, such as the one from the previous section are now
represented in a frequency form i.e.
\be
F_N(\sigma_z\otimes\sigma_z)
&=&
\frac{1}{N}\Big((\sigma_z\otimes\sigma_z)
\otimes (I\otimes I)\otimes\dots \otimes (I\otimes I)
+
\dots
+
(I\otimes I)\otimes\dots \otimes (I\otimes I)\otimes
(\sigma_z\otimes\sigma_z)\Big).\label{F_N}
\ee
One can show \cite{Hartle,FGG} that frequency operators of the form
(\ref{F_N}) naturally imply quantum mechanical laws of large numbers
(both weak and strong). 

Let us note that the sub-beam of $N$ particles $\#1$ (i.e. those
going to the ``left") is described by the product of 
reduced density matrices
\be
\rho_{\{t_0^i,t_1^i\}}(t)
&=&
\rho_{t_0^1,t_1^1}(t)
\otimes 
\dots
\otimes
\rho_{t_0^N,t_1^N}(t)\label{sub-beam}
\ee
which does not depend on times of detection of particles $\#2$. This
follows trivially from the fact that reduced
density matrices of particles $\#1$ do not depend on the form of
Hamiltonian of particles $\#2$, the fact whether the Hamiltonians are
time-dependent or not being irrelevant. The switching-off functions are an 
element
of definition of the Hamiltonians. 

Following exactly the same reasoning we can describe ensembles of 
$n$-tuples of particles and the associated $n$-time measurements. 

\section{Two-time measurements for entangled states 
in nonlinear quantum mechanics}

The above discussion does not really bring anything technically new
if one sticks to {\it linear\/} evolutions of states. 

However, if
nonlinear evolutions are involved the first two steps of our
analysis, (\ref{step1}) and (\ref{step2}), are doubtful. 
The reason is that it is not clear
whether the 
masurement performed on particle $\#1$ should be just a simple
projection producing a new initial
condition at $t=t_1$ for particle $\#2$. 
In linear quantum mechanics the projected state is an eigenvector. In
nonlinear theory there are many inequivalent definitions of
eigenvectors \cite{MC96}. Moreover, if Doebner-Goldin equations are
concerned the projection postulate should be formulated in a gauge
independent way \cite{Lucke}

Another point is that nonlinear dynamics is very
sensitive to modifications of initial conditions and in the
generic case the reduced density matrix of particle $\#2$ will depend
on the choice of both $t_1$ and the projector which performs the
reduction. The well known Gedankenexperiment of Gisin 
 \cite{Gisin} 
is precisely a two-time measurement involving steps
(\ref{step1})-(\ref{step2}) and a nonlinear evolution for
$t_1<t<t_2$. It is known to predict an unphysical behavior of the
particles and some kind of ``faster-than-light" effect. 

One should know that an appropriately (or, rather, inappropriately) 
defined nonlinear dynamics of entangled states can produce similar
unphysical phenomena 
\cite{MCu}
even if one does not explicitly use the projection postulate at the
level of solutions. The Weinberg formalism 
\cite{Weinberg}
incorporates the projection postulate already at the level of
Hamiltonian function of the composite system. In this way one defines
a multiparticle nonlinear Schr\"odinger equation whose solutions
imply probabilities one would have derived from the two step procedure
(\ref{step1})-(\ref{step2}). 

It was shown later by Polchinski 
 \cite{Polchinski}
and others 
 \cite{Jordan,MC98}
how to redefine the dynamics in a way which eliminates the unwanted
properties of the Weinberg formalism. 
To get rid of the unphysical effect associated with the projection
postulate Polchinski decided to eliminate the projection postulate
itself. But then it turned out that strange effects occur at the
level of the Many Worlds Interpretation and the ``EPR phone" is
replaced by an ``Everett phone". 
One cannot completely
disagree with those who are not yet fully satisfied by the Polchinski
formulation
\cite{Dreams,Gisin,Goldin,Mielnik00}

In the next section we propose a new framework, a sort of ``golden
middle" between Gisin and Polchinski. Although we keep the
projection postulate and do not need Many Worlds (hence no
Everett phone) we nevertheless
maintain the local propeties of the Polchinski formalism (no EPR
phone). 

\section{Switching-off-function modification of Polchinski's multiparticle
extension} 

Nonlinear dynamics in the Weinberg-Polchinski quantum mechanics is
given by Schr\"odinger equations associated with non-bilinear
Hamiltonian functions,
\be
i\dot \psi_k &=& \frac{\partial {\cal H}(\psi,\bar\psi)}{\partial \bar\psi_k}.
\ee
It is essential that not all the possible Hamiltonian functions are
acceptable. One accepts only those ${\cal H}(\psi,\bar\psi)$ which
can be written as 
\be
{\cal H}(\psi,\bar\psi)={\cal
H}(\rho)\big|_{\rho=|\psi\rangle\langle\psi|}.
\ee
For example
\be
{\cal H}(\psi,\bar\psi)
&=&
{\cal H}(\psi_+,\psi_-,\bar\psi_+,\bar\psi_-)\nonumber\\
&=& (\psi_+\bar\psi_-+\psi_-\bar\psi_+)^2
=
\langle
\psi|\sigma_x|\psi\rangle^2
=
(\Tr\rho\sigma_x)^2\big|_{\rho=|\psi\rangle\langle\psi|}
\ee
is acceptable, whereas 
\be
{\cal H}(\psi,\bar\psi)
&=&
(\psi_+\psi_-+\bar\psi_-\bar\psi_+)^2
\ee
is not (as opposed to the acceptable functions 
the latter is not invariant under $|\psi\rangle\mapsto 
e^{i\alpha}|\psi\rangle$). 
In linear quantum mechanics all Hamiltonian functions can be
written as 
\be
{\cal H}(\psi,\bar\psi)
&=&
\langle\psi|H|\psi\rangle
=
\Tr\rho H\big|_{\rho=|\psi\rangle\langle\psi|}
\ee
and, hence, are acceptable. 

If we have two such particles each described by its own Hamiltonian
function 
\be
{\cal H}_1(\psi_1,\bar\psi_1)
&=&{\cal H}_1(\rho)\big|_{\rho=|\psi_1\rangle\langle\psi_1|},\\
{\cal H}_2(\psi_2,\bar\psi_2)
&=&{\cal H}_2(\rho)\big|_{\rho=|\psi_2\rangle\langle\psi_2|},
\ee
then the two-particle Hamiltonian function is simply their sum
evaluated in appropriate one-particle states of particles $\#1$ and
$\#2$, respectively. 

What makes this formulation ingeniously simple is the fact that for a
generic {\it entangled\/} state
\be
|\Psi\rangle
&=&
\sum_{k_1k_2}\Psi_{k_1k_2}|k_1\rangle|k_2\rangle
\ee 
representing the two-particle system, states of 
the one particle subsystems may be
represented by reduced density matrices
\be
\rho_1
&=&
\sum_{k_1l_1 k_2}\bar \Psi_{k_1k_2}\Psi_{l_1k_2}
|k_1\rangle\langle l_1|,\\ 
\rho_2
&=&
\sum_{k_1 k_2 l_2}\bar \Psi_{k_1k_2}\Psi_{k_1l_2}
|k_2\rangle\langle l_2|.
\ee
Due to the above mentioned acceptability condition it makes sense to
consider 
\be
{\cal H}_{1+2}(\Psi,\bar\Psi)
&=&{\cal H}_1(\rho)\big|_{\rho_1}
+
{\cal H}_2(\rho)\big|_{\rho_2}. 
\ee
The two-particle Schr\"odinger equation has again the Hamilton form 
\be
i\dot \Psi_{k_1k_2} &=& \frac{\partial {\cal H}_{1+2}(\Psi,\bar\Psi)}
{\partial \bar\Psi_{k_1k_2}}.\label{1+2}
\ee
Having found its solution we can write with its help reduced density
matrices of the subsystems. 
It can be shown in several different ways and at different levels of
generality 
 \cite{Polchinski,Jordan,MC97,MC98} 
that the dynamics of a reduced density matrix of one of
those subsystems is
independent of the choice of Hamiltonian function of the other
subsystem. Therefore it is not posible to influence particle $\#1$ by
measurements performed on particle $\#2$, and vice versa.
This establishes locality of the formalism.

Applying the above technique to the EPR situation we have no difficulty
with describing measurements of, say, $\sigma_z\otimes\sigma_z$ if
the measurements are performed at the same time $t=t_1=t_2$. 
The formula for probabilities is the standard one:
\be
\langle \Psi(t)|E_\pm\otimes E_{\pm'}|\Psi(t)\rangle.
\ee
But what about two-time measurements? We do not have to our disposal
the projection at $t_1$ since we know that it will produce unphysical
nonlocal effects. 

What we can do, however, is to resort to the switching-off function
formulation which, as we have shown above, is equivalent to the one
based on the
projection postulate in the linear case. We define the two particle
Hamiltonian function by 
\be
{\cal H}_{t_1,t_2}(t,\Psi,\bar\Psi)
&=&\theta(t-t_1){\cal H}_1(\rho)\big|_{\rho_1}
+
\theta(t-t_2){\cal H}_2(\rho)\big|_{\rho_2}. 
\ee
The Schr\"odinger equation for two particles is again 
\be
i\dot \Psi_{k_1k_2} &=& 
\frac{\partial {\cal H}_{t_1t_2}(t,\Psi,\bar\Psi)}
{\partial \bar\Psi_{k_1k_2}}.\label{K1+2}
\ee
It is important that, similarly to the linear case, the reduced
density matrices of the subsystems depend only on their ``own"
$t_i$s. This is a straightforward consequence of locality of the
Polchinski formulation. 

\section{Example: Evolution of a pair of spin-1/2 particles}

The Hamiltonian functions in this example are 
\be
{\cal H}_1(\psi_1,\bar\psi_1)
&=&
A\langle \psi_1|\sigma_z|\psi_1\rangle^2/2\\
{\cal H}_2(\psi_2,\bar\psi_2)
&=&
B\langle \psi_2|\sigma_z|\psi_2\rangle^2/2.
\ee
The Polchinski two-particle extension is
\be
{\cal H}_{1+2}(\Psi,\bar\Psi)
&=&
A\langle \Psi|\sigma_z\otimes I|\Psi\rangle^2/2
+
B\langle \Psi|I\otimes \sigma_z|\Psi\rangle^2/2
\ee
and the two-particle Schr\"odinger equation derived from this
Hamiltonian function is 
\be
i|\dot \Psi\rangle
&=&
\Big(
A\langle \Psi|\sigma_z\otimes I|\Psi\rangle
\sigma_z\otimes I
+
B\langle \Psi|I\otimes \sigma_z|\Psi\rangle
I\otimes \sigma_z
\Big)
|\Psi\rangle.
\ee
Our modification is 
\be
{\cal H}_{t_1,t_2}(\Psi,\bar\Psi)
&=&
\theta(t-t_1)A\langle \Psi|\sigma_z\otimes I|\Psi\rangle^2/2
+
\theta(t-t_2)B\langle \Psi|I\otimes \sigma_z|\Psi\rangle^2/2
\ee
and
\be 
i|\dot \Psi\rangle
&=&
\Big(
\theta(t-t_1)A\langle \Psi|\sigma_z\otimes I|\Psi\rangle
\sigma_z\otimes I
+
\theta(t-t_2)B\langle \Psi|I\otimes \sigma_z|\Psi\rangle
I\otimes \sigma_z
\Big)
|\Psi\rangle,
\ee
with the general solution 
\be
|\Psi_{t_1,t_2}(t)\rangle
&=&
e^{-i A\langle \Psi(0)|\sigma_z\otimes I|\Psi(0)\rangle
\sigma_z\otimes I \int_0^t\theta(\tau-t_1)d\tau
-i
B\langle \Psi(0)|I\otimes \sigma_z|\Psi(0)\rangle
I\otimes \sigma_z
\int_0^t\theta(\tau-t_2)d\tau}
|\Psi(0)\rangle\nonumber\\
&=&
e^{-i A \langle\sigma_z(0)\rangle_1\sigma_z 
\kappa(t,t_1)}
\otimes
e^{-iB \langle\sigma_z(0)\rangle_2
\sigma_z 
\kappa(t,t_2)}
|\Psi(0)\rangle\label{gen-sol}
\ee
where $\langle\sigma_z(0)\rangle_k=\Tr(\rho_k(0)\sigma_z)$, 
$\kappa(t,t_k)=\int_0^t\theta(\tau-t_k)d\tau$.
(\ref{gen-sol}) describes the entire history of the two particles:
From their ``birth" at $t=0$ to their ``deaths" at $t=t_1$ and
$t=t_2$. The standard Polchinski solution is recovered in the limits
$t_1,t_2\to +\infty$. 

There is absolutely no ambiguity in the switching-off-function formulation.
An experimentalist who wants to know what are the theoretical
predictions for his experiment has simply to look into his data and
insert the detection times, $t_1$ and $t_2$, into (\ref{gen-sol}).  

An experiment with the beam consisting of $N$ pairs is described by
(\ref{N pairs}) with each of the entries of the form (\ref{gen-sol})
but with initial conditions taken at $t_0^i$s. 

To simplify further analysis we can assume
that for all the pairs the times of flight $\Delta t_k^i=t_k^i-t_0^i$,
$k=1,2$, $i=1,\dots,N$, are the same and equal $\Delta t_k$. 
Under such assumptions averages of observables, say, $X\otimes Y$,
can be computed as 
\be
\langle X\otimes Y\rangle_{\Psi,\Delta t_1,\Delta t_2}
=
\langle \Psi_{\Delta t_1,\Delta t_2}(t)|X\otimes Y
|\Psi_{\Delta t_1,\Delta t_2}(t)\rangle.
\ee
Averages of one-system observables, say $X$, are computed in the standard
way 
\be
\langle X\otimes I\rangle_{\Psi,\Delta t_1,\Delta t_2}
&=&
\langle \Psi_{\Delta t_1,\Delta t_2}(t)|X\otimes I
|\Psi_{\Delta t_1,\Delta t_2}(t)\rangle\nonumber\\
&=&
\Tr\Big(
e^{i A\langle \sigma_z(0)\rangle_1
\sigma_z\kappa(t,\Delta t_1)}
Xe^{-i A\langle \sigma_z(0)\rangle_2
\sigma_z\kappa(t,\Delta t_1)}
\rho_1(0)\Big).
\ee
The average does not depend on $\Delta t_2$. As we have
already said this is a consequence of general local properties of the
Polchinski extension. 

It is instructive to compare our result with the one we would have
obtained on the basis of standard projection-postulate computation. 
Assume that until the first measurement the evolution is the same as
before. 
At $t=t_1$ one performs a measurement of spin in some
direction $\bbox a$, i.e. the observable is 
$X=\bbox a\cdot \bbox\sigma$. The corresponding projectors are 
$E_\pm=(I\pm X)/2$. Again we divide the
calculation into steps.
\begin{itemize}
\item
At $t=t_1$ the state is
\be
|\Psi(t_1)\rangle
&=&
e^{-i A \langle\sigma_z(0)\rangle_1\sigma_z  t_1}
\otimes
e^{-iB \langle\sigma_z(0)\rangle_2
\sigma_z  t_1}
|\Psi(0)\rangle.
\ee
\item
Using the Zeno-type
argument we reduce at $t=t_1$ by 
\be
|\Psi(t_1)\rangle
\mapsto
\frac{E_\pm\otimes I |\Psi(t_1)\rangle}
{\parallel E_\pm\otimes I |\Psi(t_1)\rangle \parallel}
=:|\Psi_\pm(t_1).\rangle\label{psi_pm}
\ee
\item
We allow the ``right" particle to evolve for $t_1<t<t_2$ but starting
at $t_1$ with the initial condition (\ref{psi_pm})
\be
|\Psi_\pm(t_2)\rangle
&=&
I
\otimes
e^{-iB \langle\Psi_\pm(t_1)|I\otimes \sigma_z|\Psi_\pm(t_1)\rangle
\sigma_z  (t_2-t_1)}
|\Psi_\pm(t_1)\rangle.
\ee
\item
Compute the conditional probability 
\be
\frac{\langle \Psi(t_1)|
E_\pm
\otimes
e^{iB \langle\Psi_\pm(t_1)|I\otimes \sigma_z|\Psi_\pm(t_1)\rangle
\sigma_z  (t_2-t_1)}
E_{\pm'}e^{-iB \langle\Psi_\pm(t_1)|I\otimes \sigma_z|\Psi_\pm(t_1)\rangle
\sigma_z  (t_2-t_1)}
|\Psi(t_1)\rangle}
{\langle \Psi(t_1)|E_\pm \otimes I |\Psi(t_1)\rangle}.\label{condit'}
\ee
\item
The numerator of (\ref{condit'}) is the joint probability we are
looking for:
\be
\langle \tilde\Psi^\pm_{t_1,t_2}(t_2)|
E_\pm
\otimes
E_{\pm'}
|\tilde\Psi^\pm_{t_1,t_2}(t_2)\rangle
\ee
where 
\be
|\tilde\Psi^\pm_{t_1,t_2}(t_2)\rangle
=
e^{-iA \langle\Psi(0)|\sigma_z\otimes I|\Psi(0)\rangle \sigma_z t_1}
\otimes
e^{-iB \langle\Psi_\pm(t_1)|I\otimes \sigma_z|\Psi_\pm(t_1)\rangle
\sigma_z  (t_2-t_1)}
e^{-iB \langle\Psi(0)|I\otimes \sigma_z|\Psi(0)\rangle \sigma_z t_1}
|\Psi(0)\rangle
\ee
\end{itemize}
The latter formula can be compared with the one we have derived on
the basis of our formalism and which can be written as 
\be
|\Psi_{t_1,t_2}(t_2)\rangle
=
e^{-iA \langle\Psi(0)|\sigma_z\otimes I|\Psi(0)\rangle \sigma_z t_1}
\otimes
e^{-iB \langle\Psi(0)|I\otimes \sigma_z|\Psi(0)\rangle
\sigma_z  (t_2-t_1)}
e^{-iB \langle\Psi(0)|I\otimes \sigma_z|\Psi(0)\rangle \sigma_z t_1}
|\Psi(0)\rangle.
\ee
Figs. 2 and 3 show averages of 
$\sigma_x \otimes \sigma_x$ (solid), 
$\sigma_x\otimes I$ (dashed), and 
$I\otimes \sigma_x$ (dotted)
calculated by means of the two procedures. The initial state is 
\be
|\Psi(0)\rangle
&=&
\frac{1}{3}|1\rangle|2\rangle
-
\frac{2\sqrt{2}}{3}|2\rangle|1\rangle
\ee
where 
\be
|1\rangle
=
\left(
\begin{array}{c}
\cos(\pi/8)\\
\sin(\pi/8)
\end{array}
\right),
\quad
|2\rangle
=
\left(
\begin{array}{c}
-\sin(\pi/8)\\
\cos(\pi/8)
\end{array}
\right).
\ee
The parameters in Hamiltonians are $A=8$, $B=1/2$, and the detection
times are $t_1=3.5$ and $t_2=8$ (all in dimensionless units). 
In Fig.2 we have used the switching-function approach. The dotted
line representing the average of $I\otimes \sigma_x$ does not
``notice" the measurement performed on particle $\#1$. In Fig. 3
one can observe a slight change in the doted curve at $t=t_1$. This
is the nonlocal effect of the type described by Gisin. Until $t=t_1$
the evolution is described in the Polchinski way. As one can see the
Zeno-type reasoning leads even in this case to the nonlocal influence
between the two particles. The switching-function-modified Polchinski
description is free from the difficulty. 

\section{Pre-selection, post-selection, and teleportation}

It seems we should finally make it clear how does our procedure affect the
issue of teleportation of quantum states. Teleportation is
precisely an experiment of the type we have discussed: There are
correlated systems and information on results of measurements on one
subsystem is used to create at later times an appropriate state of the
other one. We claim there is no actual difficulty but one has to be careful
with identification of pre- and post-selected ensembles if nonlinear
evolution is involved. 

The simplest example of quantum teleportation is the following
procedure. 
\begin{itemize}
\item
A source produces a pair of particles in a singlet state.
\item
An observer $\#1$ (``Alice") performs a measurement of spin of
particle $\#1$. 
\item
If the result is $+1/2$ she phones to ``Bob" and instructs him to
remove particle $\#2$.
\item
If the result is $-1/2$ she phones to ``Bob",
and tells him to use his particle for further measurements.
\item
As a result Bob obtains a beam of particles with spin $+1/2$. 
\end{itemize}
In more sophisticated versions of teleportation one can use several
particles and measurements on both sides may be accompanied by
various additional operations. What is common to all the
teleportation schemes is the presence of the ``classical channel" (a
phone, say) which allows Alice and Bob to exchange information and
instructions. 

The above procedure should be contrasted with the {\it correlation
experiment\/} in which Alice and Bob simply perform measurements and
{\it afterwards\/}, when the experiment is completed, reject all
pairs of data where spin $+1/2$ was found for particles $\#1$. 
After having rejected a part of the data what is left may be regarded
as an experiment performed by Bob on the sub-beam of particles $\#2$
whose initial spin was $+1/2$. 

The teleportation scheme involves the so called pre-selection (Bob
deals with a partial, or pre-selected, ensemble). 
Correlation experiment is based on post-selection (Bob first deals
with the entire ensemble and {\it post factum\/} 
chooses the interesting part of the data). 

Linearity of quantum mechanics allows one to formally identify the
two procedures in spite of the fact that they involve completely
different experimental arrangements. To see the differences one
encounters in nonlinear situation consider the nonlinear
Schr\"odinger equation of the type we have used in the example. 

Let the nonlinear Hamiltonian of particle $\#1$ from our pair be
\be
A \langle \Psi|\bbox \sigma \otimes I|\Psi\rangle 
\cdot\bbox \sigma \otimes I
\sim \bbox B(\rho_1)\cdot \bbox \sigma\otimes I
\ee
where $\rho_1$ is the reduced density matrix of particle $\#1$ and 
we have introduced the magnetic field 
\be
\bbox B(\rho_1)\sim \Tr\rho_1\bbox \sigma.
\ee
which is proportional to the average magnetic moment of the beam. 
The Hamiltonian represents a mean-field interaction of a single particle
with the magnetic field created by the entire beam. 

In the singlet case the average magnetic field of the {\it entire\/}
ensemble of particles $\#1$ is zero. Alice may or may not perform her
measurements but this is irreleveant unless she instructs Bob to
undertake concrete filtering operations. However, if
Bob pre-selects a half of the ensemble then obviously the ensemble
will produce a nonvanishing magnetic field. The data collected with
pre- and post-selected ensembles will be different.
Each time Bob filters out a particle on the basis of Alice's
information, {\it he\/} creates a new initial condition. 
Preparation of initial conditions must be accompanied by a
projection. 

\section{Discussion}

The procedure of computing multiple-time correlation functions we
have introduced is not intended as a solution of the ``measurement
problem". Moreover, there are other limitations of the procedure,
especially arising in relativistic contexts. To give an example, it
is not clear how to parametrize the switching-off functions for
relativistic equations. However, from the very beginning our approach
was not meant as a fundamental one. We have concentrated only on the
part of the dynamics allowing the energy to leak out to detectors. 
In this respect we are doing neither worse not better than all
the other approaches, both quantum and classical, 
that allow time dependent Hamiltonians. 
What we have tried to do was to clarify an important thinking error
which seems to have plagued the literature on nonlinear Schr\"odinger
equations. 

There are many dangers one encounters while trying to extend linear
quantum mechanics to nonlinear domains. The reason is that linear
theories are, in a sense, very pathological. One can imagine the
difficulties we would have in classical mechanics if the only
potentials experimentally observed in Nature were those of harmonic
oscillators. No doubt there would be some ``impossibility theorems"
about nonexistence of other potentials. 

This is precisely what happens in linear quantum mechanics since
linear Schr\"odinger equation is mathematically equivalent to a
classical Hamiltonian system describing uncountably many linear
harmonic oscillators. Nonlinear Schr\"odinger equations correspond to
non-quadratic Hamiltonian functions. 

Still, there is no way of stopping people use nonlinear
Schr\"odinger equations. They are simply too useful and too
interesting, and often very natural from a mathematical point of view. 
However, once one realizes that it may be justifed to contemplate
nonlinearly evolving states, one immediately encounters another
logical trap: A generic quantum state is not represented by a ray or
a vector from a Hilbert space, but by a reduced density matrix which
is not a one-dimensional projector. 
States represented by rays are a rarity
enjoyed by completely isolated systems \cite{Mermin}. This is one of
the most fundamental consequences of the existence of entangled states
which, as we all know, are what makes quantum mechanics so special. 

It seems therefore that the psychological difficulty we have to face is that
once we accept a possibility of nonlinearly evolving states, we have
to immediately forget about any fundamental importance of nonlinear
Schr\"odinger equations. This was realized by Polchinski and this is
his great contribution to the subject. One has to start with
nonlinear generators of evolution defined on the set of generic
states, that is, density operators. But then the fundamental level of
description must be the von Neumann one. 

There is nothing wrong with such a viewpoint. The set of solutions of
nonlinear von Neumann equations is a Poisson manifold and, hence, a
classical phase space with a Lie-Poisson dynamics. In this respect
nonlinear von Neumann equations share many similarities with
rigid-body or hydrodynamical equations, and even with the Nahm
equations from Yang-Mills theories \cite{NUMC00}. Solutions of von Neumann
equations are pure states (in the classical dynamical sense) on a
Poisson manifold \cite{Bona} in the same sense as solutions of nonlinear
Schr\"odinger equations are pure states on a K\"ahler manifold
\cite{Kibble,Cirelli,Ashtekar,Hughston}. Similarly to nonlinear
Schr\"odinger equations the von Neumann ones (at least a
surprisingly rich class of them) are integrable by Lax-pair and
Darboux methods \cite{SLMC98,NUMCMKSL00}. 

It is clear that a density operator which solves a nonlinear von
Neumann equation has the same ontological status as a state-vector
solution of a Schr\"odinger equation. In our approach the density
operator represents a state of a single quantum system and not of the
entire ensemble. The ensemble is represented by a (finite or
infinite) tensor product of single-particle states. In this respect
we are not really very original since several other authors gave in
different contexts (weak measurements \cite{AA}, quantum logic
\cite{Aerts}) arguments for the same status of reduced density
matrices (derived via reduction from entangled states) 
and state vectors. 

One dimensional projectors may be regarded as forming a kind of
boundary of the Poisson manifold of pure states. To define a flow on
this manifold it is not enough to define Hamiltonian functions on the
boundary. Such a possibility indeed exists in linear quantum
mechanics, but this is an example of pathological properties of
linear Hamiltonian systems. But of course, once one has the dynamics
on the entire manifold, one can contemplate its restriction to the
boundary. And then one arrives again at nonlinear Schr\"odinger
equations although somewhat stripped of a fundamental importance. 
This is actually what we have done in the examples discussed in this
paper. 

\acknowledgments

The viewpoint on structure of nonlinear quantum mechanics we tried to
present in this paper was
gradually formed during the past ten years in discussions with D.
Aerts, I. Bia{\l}ynicki-Birula, P. B\'ona,  
N. Gisin, G. A. Goldin, J. Hoenig, T. F. Jordan, M. Kuna, S. Leble,
W. L\"ucke, W. A. Majewski, M. Marciniak, B. Mielnik, P. Nattermann, J.
Naudts, A. Posiewnik, K. Rz\c a\.zewski, and {\L}. A. Turski. 

MC thanks Alexander von Humboldt Foundation for making possible his
stay in Clausthal where this work was done, Polish Committee for
Scientific Research for support by means of the KBN Grant No. 5 PO3B 040 20,
and Ania and Wojtek Pytel for the notebook computer used to type-in
this paper.

\section{Appendix: Nonlinear averages}

We define and use averages in the standard way: Averaging is linear
and random variables are represented by Hermitian operators. In this
way we avoid the arguments given by Mielnik \cite{Mielnik00}. But
Hamiltonian functions are not represented by linear averages,
otherwise the evolution would be linear. So to seriously face
Mielnik's objections we have to say something about probability
interpretation of nonlinear observables.

We are aware of at least two situations from a purely classical
domain where several different ways of averaging (linear and
nonlinear) may coexist simultaneously. What seems important,
similarly to Hamiltonian equations of motion, they are
related to optimization (i.e. variational) problems. 
The first class of examples is due to R\'enyi
and is associated with his $\alpha$-entropies. 

\subsection{Kolmogorov-Nagumo averages and R\'enyi entropies}

Let $E$ be the disjoint union of the sets
$E_1,\dots, E_n$ having $N_1,\dots,N_n$ elements respectively
$\bigl(\sum_{k=1}^{n} N_k=N\bigr)$. Let us suppose that we are
interested only in knowing the subset $E_k$. 
The information characterizing an
element of $E$ consists of two parts: The first specifies the
subset $E_k$ containing this particular element and the second
locates it within $E_k$. The amount of the second piece of
information is, by Hartley formula, $\log_aN_k$ \cite{Hartley}. 
On the other hand, to specify an element of $E$
we need $\log_aN$ units of information. The amount necessary for
the specification of the set $E_k$ is therefore 
\be
I_k=\log_aN -
\log_aN_k=\log_a\frac{N}{N_k}=\log_a\frac{1}{p_k}. 
\ee
It follows that the amount of information received by
learning that a single
event of probability $p$ took place equals
\be
I(p)=\log_a\frac{1}{p}.
\ee
This {\it random variable\/} is very interesting and unusual. It can be
regarded as a classical example of a {\it nonlinear observable\/}.  
 
In statistical situations  measured quantities correspond
to averages of random variables. Therefore the average
information is 
\be
I=\sum_k p_k\log_a\frac{1}{p_k}.\label{Shannon}
\ee
This is the Shannon formula and $I$ is called the {\it
entropy\/} of the probability distribution $\{p_1,\dots,p_n\}$ \cite{Shannon}.

R\'enyi gave examples of information theoretic questions
where measures of information are those obtained by more
general ways of averaging --- the Kolmogorov--Nagumo (KN) function
approach \cite{Renyi}. 
Let $\varphi$ be a monotonic function.
The KN average information can be defined by
means of 
$\varphi$ as 
\be
I=\varphi^{-1}\Biggl(\sum_k
p_k\varphi\Bigl(\log_a\frac{1}{p_k}\Bigr)\Biggr). 
\ee
If the generalized information measure is to satisfy the
postulate of additivity, the function $\varphi$ has to be 
linear or exponential (up to an affine transformation $\varphi\mapsto
a \varphi+b$ which leaves all KN averages invariant).
The linear function corresponds to Shannon's
information. Choosing
$\varphi(x)=a^{(1-\alpha)x}$ we find
\be
I_\alpha^R=\varphi^{-1}\Biggl(\sum_k
p_k\varphi\Bigl(\log_a\frac{1}{p_k}\Bigr)\Biggr)=
\frac{1}{1-\alpha}\log_a\Bigl(\sum_kp_k^\alpha\Bigr).
\label{alpha}
\ee
Formula (\ref{alpha}) describes R\'enyi's $\alpha$-entropy.

The nonlinear Hamiltonian functions we have used in the
example are of the form 
\be
{\cal H}&\sim& (p_+-p_-)^2\nonumber\\
&=&\varphi^{-1}\big(\varphi(9)p_++\varphi(1)p_-\big)
-4(p_+-p_-)-4,
\ee
with $p_\pm=|\psi_\pm|^2$, $\varphi(x)=\sqrt{x}$. It follows that 
the Hamiltonian functions consist of two averages, one linear and one
of the KN type. 

For general density matrices KN averages are 
\be
\langle X\otimes Y\rangle_{KN}
&=&
\varphi^{-1}\big(\Tr \rho \varphi(X\otimes Y)\big)\\
\langle X\otimes I\rangle_{KN}
&=&
\varphi^{-1}\big(\Tr \rho \varphi(X\otimes I)\big)\\
&=&
\varphi^{-1}\big(\Tr \rho \varphi(X)\otimes I\big)\\
&=&
\varphi^{-1}\big(\Tr_1 \rho_1 \varphi(X)\big)
\ee
the latter being consistent with the Polchinski definition ($\Tr_1$
is the partial trace over the subsystem).

\subsection{$q$-averages and Tsallis entropies}

Tsallis non-extensive thermodynamics \cite{T88,TMP98} 
is based on the entropy 
\be
S_q=\frac{\sum_kp_k^q-1}{1-q}
\ee
and internal energy 
\be
H_q=\frac{\sum_kp_k^q E_k}{\sum_kp_k^q}=:\sum_kP^{(q)}_k E_k.
\ee
Thermodynamic equilibria are calculated by the standard variational
principle which uses $q$-free energy 
$F_q=H_q-TS_q$ 
and the constraint $\sum_kp_k=1$. Therefore there are two sets of
probabilities in such a theory, $p_k$ and $P^{(q)}_k$, both
normalized to unity. 
 
The definition of entropy becomes more natural if one recalls the formula
\be
p^q-p=pe^{(q-1)\ln p}-p\approx (q-1)p\ln p.
\ee
This also sheds some light on the meaning of the parameter $q$ in 
Tsallis' definition: $q$ measures deviations from exponential and
logarithmic functions. Thinking of the Shannon derivation of entropy
we have given in the previous subsection one can see that $q$ has
something to do with possibilities of dividing ensembles into
subensembles. That this is the case can be seen also in the formula
for $q$-entropy of two independent systems ($A$ and $B$)
\be
S_q(A\cup B)=
S_q(A)+S_q(B)+(1-q)S_q(A)S_q(B).\label{AuB}
\ee
Tsallis statistics has a wide range of applications. There are
physical systems that are well described by $q$'s quite far from
unity. In plasma one finds $q\approx 0.5$, in multiparticle
production process (hadronization) the best fit of experimental 
data yields $q\approx 0.72$. $q$-statistics arises whenever in the
system one encounters long-range correlations, memory effects or
fractal structures. 

In a sense even more interesting from our perspective are purely
classical applications of $q$-statistics since they provide many
insights into the meaning of non-linear averages. Among such
applications one finds statistics of goals in football championships,
citations o scientific papers, the Zipf-Mandelbrot law in linguistics
and sociology, and many others.

From the point of view of nonlinear quantum dynamics it is
interesting to look at neighborhoods of Tsallis {\it
thermodynamical\/} equilibria. A
variational principle based on $q$-averaged energy leads to the
nonlinear von Neumann equation \cite{MCJN99}
\be
i\dot\rho\sim [H,\rho^q]
\ee
and free energy $F_q$ plays a role of a stability function which
characterizes a {\it dynamical\/} equilibrium.

\twocolumn

\begin{figure}
\epsfxsize=8.25cm
\epsffile{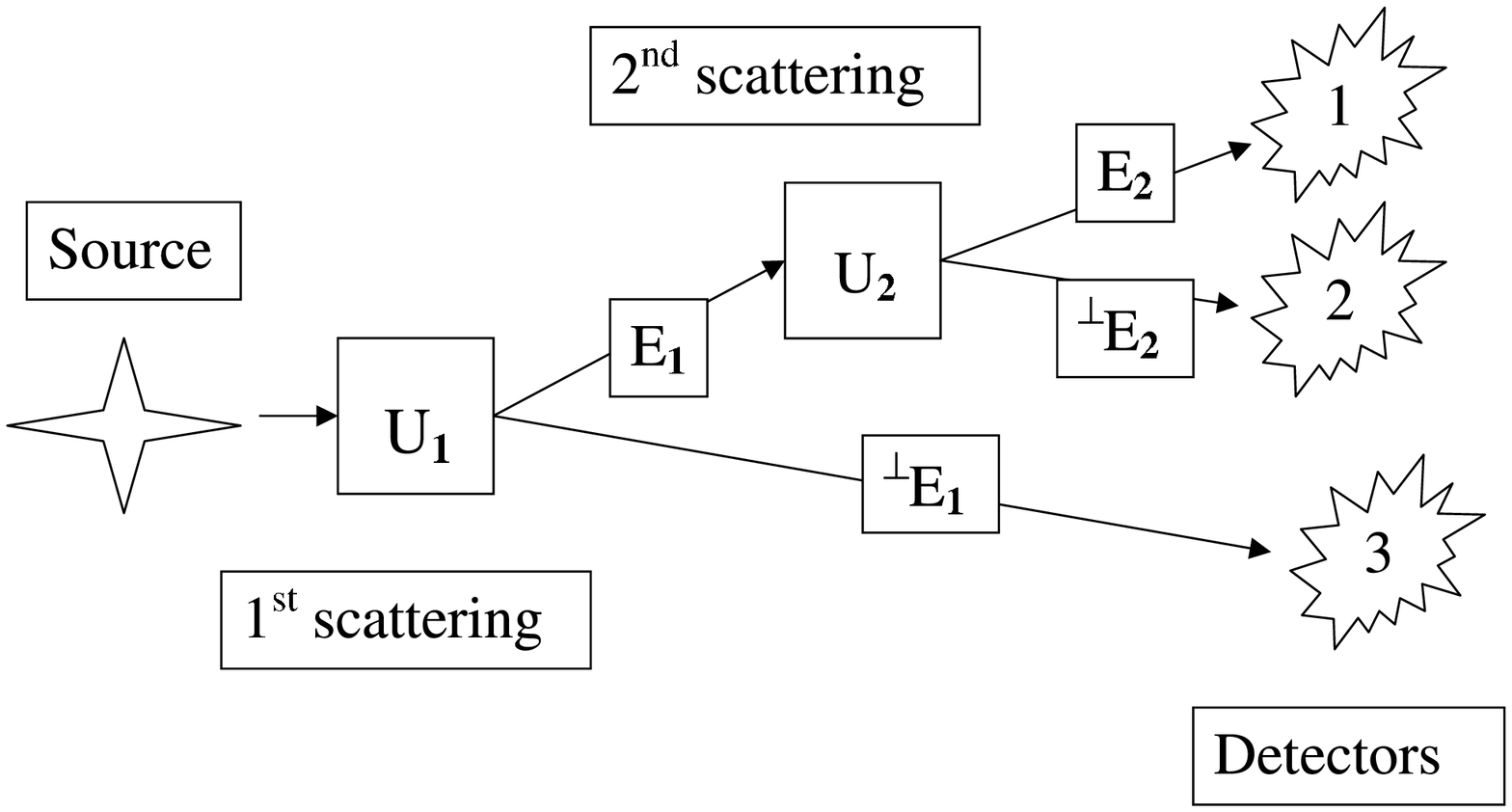}
\caption{A history. Projector $E_1$ does not ``reduce" states. 
Filters $U_1$ and $U_2$ are represented by unitary operators.}
\end{figure}
\begin{figure}
\epsfxsize=8.25cm
\epsffile{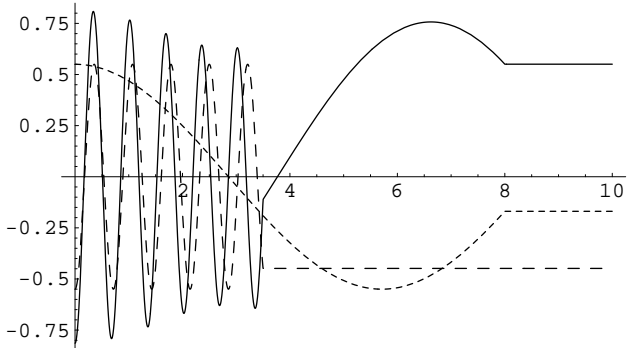}
\caption{Averages of the three observables in the switching-function
formulation. The dotted line shows the evolution of observable
$\sigma_x$ associated with particle $\#2$ which is detected at
$t=t_2=8$. Earlier detection of particle $\#1$ at $t_1=3.5$ does not
influence particle $\#2$.} 
\end{figure}
\begin{figure}
\epsfxsize=8.25cm
\epsffile{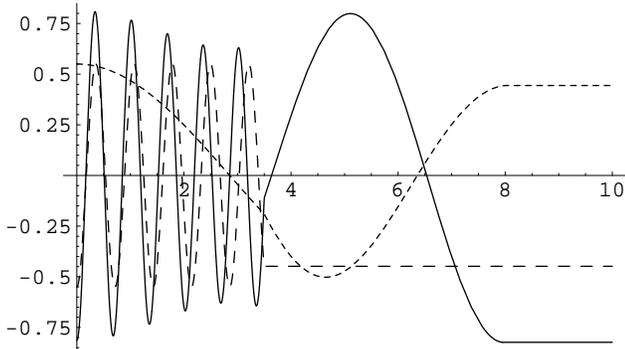}
\caption{Averages of the three observables calculated in a Zeno-type
way. Measurement at $t=t_1=3.5$ performed on particle $\#1$
nonlocally influences the behavior of particle $\#2$. As opposed to
the plot from Fig. 2 the dotted line is modified at $t=3.5$. This is
Gisin-type nonlocality.} 
\end{figure}

\begin{references}
\bibitem{Bona1}P. B\'ona discussed this in detail in \cite{Bona}. He
distinguishes between ``elementary mixtures" (in our terminology
these are pure states on the Poisson manifold) and ``barycenters"
which are mixed in the standard sense. The main point is that
generic states of subsystems are represented by reduced density
matrices. The fact that they are not one dimensional projectors has
an entirely quantum origin and has nothing to do with the so called
``ignorance interpretation of quantum ensembles". Similar oppinions
on the ontological status of reduced density matrices 
were expressed in \cite{AA,Mermin,Aerts}.
\end{references}
\end{document}